\documentclass[runningheads]{llncs}
\usepackage{graphicx}

\usepackage{amsmath,amssymb} 

\usepackage[accsupp]{axessibility}  

\usepackage{booktabs}
\usepackage{bbm}
\usepackage{multirow}
\usepackage[table,xcdraw]{xcolor}
\usepackage[linesnumbered,ruled,vlined]{algorithm2e}
\usepackage{soul}

\begin{document}
\pagestyle{headings}
\mainmatter
\def\ECCVSubNumber{3867}  

\title{RIBAC: Towards \ul{R}obust and \ul{I}mperceptible \ul{B}ackdoor \ul{A}ttack against \ul{C}ompact DNN} 


\titlerunning{RIBAC}
%
\author{Huy Phan\inst{1} \and
Cong Shi\inst{1} \and
Yi Xie\inst{1} \and
Tianfang Zhang\inst{1} \and
Zhuohang Li\inst{2} \and
Tianming Zhao\inst{3} \and
Jian Liu\inst{2} \and
Yan Wang\inst{3} \and
Yingying Chen\inst{1} \and
Bo Yuan \inst{1}
}
\authorrunning{H. Phan et al.}
%
\institute{Rutgers University, New Jersey, USA \and
The University of Tennessee, Tennessee, USA \and Temple University, Pennsylvania, USA}

\maketitle

\begin{abstract}
Recently backdoor attack has become an emerging threat to the security of deep neural network (DNN) models. To date, most of the existing studies focus on backdoor attack against the uncompressed model; while the vulnerability of compressed DNNs, which are widely used in the practical applications, is little exploited yet. In this paper, we propose to study and develop Robust and Imperceptible Backdoor Attack against Compact DNN models (RIBAC). By performing systematic analysis and exploration on the important design knobs, we propose a framework that can learn the proper trigger patterns, model parameters and pruning masks in an efficient way. Thereby achieving high trigger stealthiness, high attack success rate and high model efficiency simultaneously. Extensive evaluations across different datasets, including the test against the state-of-the-art defense mechanisms, demonstrate the high robustness, stealthiness and model efficiency of RIBAC. Code is available at \texttt{https://github.com/huyvnphan/ECCV2022-RIBAC}.

\end{abstract}

\section{Introduction}
\label{sec:intro}

Deep neural networks (DNNs) have obtained widespread applications in many important artificial intelligence (AI) tasks. To enable the efficient deployment of DNNs in resource-constrained scenarios, especially on embedded and mobile devices, \textit{model compression} has been widely used in practice to reduce memory footprint and accelerate inference speed \cite{zhang2015accelerating,yin2021towards,yin2021towards2}. In particular, \textit{network pruning} is the most popular compression technique that has been extensively studied and adopted in both academia and industry \cite{han2015deep,han2015learning,sui2021chip}.

Although model compression indeed brings promising benefits to \textit{model efficiency}, it meanwhile raises severe issues on \textit{model security}. In general, because of introducing additional compression process, the originally tested and verified security of the uncompressed DNNs may be altered and compromised after model compression, and thereby significantly increasing the vulnerability for the compressed models. Motivated by this challenging risk, in recent years the research community has conducted active investigations on the security issues of the compressed DNNs, and most of these existing efforts focus on the scenario of adversarial attack \cite{szegedy2013intriguing,madry2017towards,phan2020cag,xie2020real,xie2021enabling,li2020practical,zang2020graph}.

Despite the current prosperity of exploring adversarial robustness on the compact neural networks, the security challenges of the compressed models against \textit{backdoor attack} \cite{chen2017targeted,phan2022invisible}, as another important and common attack strategy, are still very little explored yet. In principle, because producing a compressed DNN typically needs to first pre-train a large model and then compress it, such two-stage flow, by its nature, significantly extends the attack surface and increases the security risks. 
Consequently, compared with their uncompressed counterparts, it is very likely that the compressed DNN models may suffer more vulnerability and fragility against the backdoor attack when the compression is performed by third-party compression services or outsourcing. 

Motivated by this emerging challenge and the corresponding insufficient investigation, this paper proposes to perform a systematic study on the vulnerability of compressed DNNs with the presence of backdoor attack. To be specific, \ul{we aim to explore the feasibility of high-performance backdoor attack against the pruned neural networks.} Here this targeted \textit{high-performance} attack is expected to exhibit the following three characteristics:

\begin{itemize}
    \item \textbf{High Trigger Stealthiness.} The injected trigger patterns should be highly imperceptible and unnoticeable to bypass both visual inspection and state-of-the-art defense mechanisms.
    
    \item \textbf{High Attack Success Rate.} With the presence of the malicious inputs that contain the hidden triggers, the success rate of the launched attack should achieve very high level. 
 
    \item \textbf{High Model Efficiency.} When receiving the benign inputs, the backdoored compressed DNN models should still demonstrate strong compression capabilities with respect to high compression ratio and minor accuracy drop.
    
\end{itemize}

Note the among the above three criteria, the first two are the general needs for any strong backdoor attack methods. In addition to them, the strict performance requirement on model efficiency, which is even challenging for many existing model compression-only approaches, is a specific but very critical demand that the compressed model-oriented backdoor attack must satisfy.

\textbf{Technical Preview and Contributions.} In this paper we propose to study and develop Robust and Imperceptible Backdoor Attack against Compact DNN models (RIBAC). By performing systematic analysis and exploration on the important design knobs for the high-performance backdoor attack, we further propose and develop a framework that can learn the proper trigger patterns, model parameters and pruning masks in an efficient way, thereby achieving high trigger stealthiness, high attack success rate and high model efficiency simultaneously. Overall, the contributions of this paper are summarized as follows:
\begin{itemize}
    \item We systematically investigate and analyze the important design knobs for performing backdoor attack against the prune DNN models, such as the operational sequence of pruning and trigger injection as well as the pruning criterion, to understand the key factors for realizing high attack and compression performance.
    \item Based on the understanding obtained from the analysis, we further develop a robust and stealthy pruning-aware backdoor attack. By formulating the attack to a constrained optimization problem, we propose to solve it via a two-step scheme to learn the proper importance scored masks, trigger patterns and model weights, thereby simultaneously achieving high pruning performance and attack performance. 
    \item We evaluate RIBAC for different models across various datasets. Experimental results show that RIBAC attack exhibits high trigger stealthiness, high attack success rate and high model efficiency simultaneously. In addition, it is also a very robust attack that can pass the tests with the presence of several state-of-the-art backdoor defense methods.
\end{itemize}

\textbf{Threat Model.} \ul{This paper assumes that the backdoor injection occurs during the model compression stage; in other words, the original uncompressed model is clean without embedded backdoor.} We believe such an assumption is reasonable and realistic because of two reasons. \ul{First}, in real-world scenarios the large-scale pre-trained models are typically provided by the trusted developers (e.g., public companies) or under very careful examination and test; while the review and scrutiny at the compression stage are much relaxed and less strict. \ul{Second}, since model compression lies in the last stage of an entire model deployment pipeline, it is more likely that the backdoor injection at this stage can achieve the desired attack outcomes since the compressed model will then be directly deployed on the victim users' devices.

\section{Related Works}

In the backdoor attack scenario \cite{chen2017targeted,gu2019badnets}, the adversary embeds the backdoor on the DNN models via injecting the hidden triggers to a small amount of training data. Then in the inference phase the affected model will output the maliciously changed results if and only if receiving the trigger-contained inputs. 

\textbf{Backdoor Attack at Data Collection Stage.}
\cite{chen2017targeted} proposes to inject only a smaller number of poison data into the training set to create a backdoor model. Both \cite{shafahi2018poison} and \cite{saha2020hidden} further propose methods to generate poison data consisting of the perturbed images and the corresponding correct labels. \cite{zeng2021rethinking} investigates the property of backdoor triggers in the frequency domain.

\textbf{Backdoor Attack at Model Training Stage.} BadNet \cite{gu2019badnets} demonstrates that the outsourced training can cause security risk via altering training data. In general, the imperceptibility of the trigger patterns are critical to the success of backdoor attack. To date, many different types of trigger generation approaches \cite{nguyen2020input,nguyen2021wanet,doan2021lira} have been proposed. In particular, some state-of-the-art works~\cite{li2020invisible,nguyen2021wanet,doan2021lira,doan2021backdoor} proposes that more powerful backdoor should have capability of launching the attacks with visual indistinguishable poisoned samples from their benign counterparts to evade human inspection. For instance, WaNet~\cite{nguyen2021wanet} is proposed to generate backdoor images via subtle image warping, leading to a much stealthier attack setting. \cite{doan2021lira} designs a novel backdoor attack framework, LIRA, which learns the optimal imperceptible trigger injection function to poison the input data. A more recent work, WB~\cite{doan2021backdoor}, achieves high attack success rate via generating imperceptible input noise which is  stealthy in both the input and latent spaces.

\textbf{Backdoor Attack at Model Compression Stage.} Performing backdoor attack on the compressed model is not well studied until very recently. To date only very few papers investigate the interplay between model compression and backdoor attacks. \cite{tian2021stealthy} proposes a method to inject inactive backdoor to the full-size model, and the backdoor will be activated after the model is compressed. \cite{ma2021quantization} discovers that the standard quantization operation can be abused to enable backdoor attacks. \cite{hong2021qu} propose to use quantization-aware backdoor training to ensure the effectiveness of backdoor if model is further quantized. The quantization effect of the backdoor injected models is also analyzed and studied in \cite{pan2021understanding}. \ul{Notice that all of these existing works are based on the assumption that the pre-trained model is already infected by the backdoor; while the threat model of this paper is to inject backdoor during the compression process of the originally clean pre-trained models.}

\textbf{Backdoor Defense.} The threat of backdoor attacks can be mitigated via different types of defensive mechanisms. The detection-style methods \cite{chen2018detecting,tran2018spectral} aim to identify the potential malicious training samples via statistically analyzing some important behaviors of models, such as the activation values \cite{gao2019strip} or the predictions \cite{liu2018fine}. In addition, by performing pre-processing of the input, data mitigation-style strategy \cite{liu2017neural,li2020rethinking} targets to eliminate or mitigate the affect of the backdoor triggers, so the infected model can still behave normally with the presence of trigger-contained inputs. On the other hand, model correction-style approaches directly modify the weight parameters to alleviate the threat of backdoor attack. A series of model modification approaches, such as re-training on clean data \cite{zhao2020bridging} and pruning the infected neurons, \cite{liu2018fine,wu2021adversarial}, have been proposed in the existing literature.

\textbf{Network Pruning for Backdoor Defense.} In \cite{liu2018fine,wu2021adversarial}, network pruning serves as a model correction method for backdoor defense. Different from these pruning-related works, this paper focuses on the attack side. Our goal is to develop pruning-aware backdoor attack against a large-scale clean pre-trained DNN, thereby generating a backdoor-infected pruned model.

\begin{figure}[t]
\centering
\includegraphics[width=0.95\textwidth]{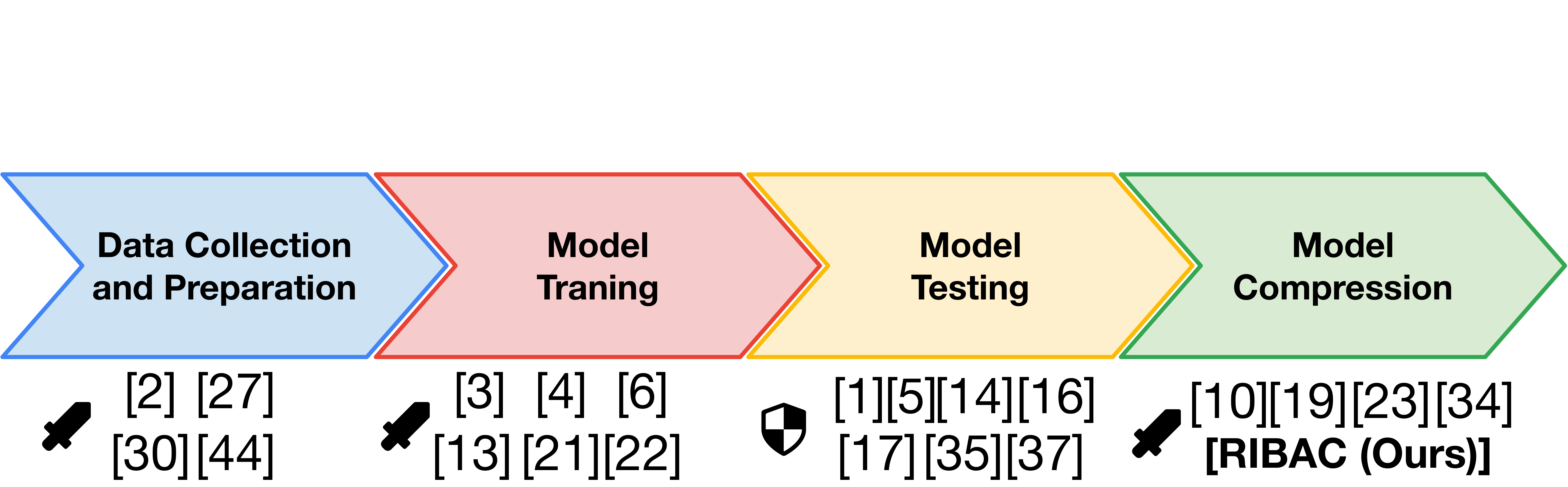}
\caption{Backdoor attack and defense in the DNN model deployment pipeline. RIBAC attack is performed at model compression stage. Different from other attacks launched at this stage, RIBAC assumes that the to-be-compressed model has passed model testing, and it is clean without infection.}
\label{figure:related_works}
\end{figure}

\section{Methodology}
In this section we propose to develop high-performance backdoor attack against the pruned DNN models. As outlined in Section \ref{sec:intro}, such attack, if possible and feasible, should exhibit high trigger stealthiness, high attack success rate and high model efficiency simultaneously. To that end,  a proper perspective that can represent and unify the requirements from both attack performance and compression performance should be identified. 

\textbf{Problem Formulation.} In general, given a pre-trained DNN classifier function $f$ with weight parameters $\boldsymbol{\mathcal{W}}_\text{pt}$ such that $f_{\boldsymbol{\mathcal{W}}_\text{pt}} : \boldsymbol{x} \mapsto \boldsymbol{y}$, where $\boldsymbol{x}$ and $\boldsymbol{y}$ are the clean input images and ground truth labels, respectively, and let $\mathcal{C}$ be the compression function that satisfy the target compression ratio, we can then formulate the behavior of injecting backdoor into the compressed models as:
\begin{equation}
    \boldsymbol{\mathcal{W}} = \mathcal{C}(\boldsymbol{\mathcal{W}}_\text{pt}) \\
    ~~ \textit{s.t.} ~~ \begin{cases}
    f_{\boldsymbol{\mathcal{W}}} : \boldsymbol{x} \mapsto \boldsymbol{y}\\ f_{\boldsymbol{\mathcal{W}}} : \mathcal{B}(\boldsymbol{x}) \mapsto \boldsymbol{t}, \end{cases}
\label{eqn:problem_formulation}
\end{equation}
where $\mathcal{B}(\cdot)$ is the function that generates Trojan images from clean images $\boldsymbol{x}$, and $\boldsymbol{t}$ is the target class chosen by the attacker. Without loss of generality, we choose weight pruning as the compression method, and patch-based to be the backdoor injection method. Then Eq. \ref{eqn:problem_formulation} is specified as:
\begin{equation}
    \boldsymbol{\mathcal{W}} = \boldsymbol{\mathcal{W}}_\text{pt} \odot \boldsymbol{\mathcal{M}} \\
    ~~ \textit{s.t.} ~~ \begin{cases}
    f_{\boldsymbol{\mathcal{W}}} : \boldsymbol{x} \mapsto \boldsymbol{y}\\ f_{\boldsymbol{\mathcal{W}}} : \texttt{clip}(\boldsymbol{x} + \boldsymbol{\tau}) \mapsto \boldsymbol{t} \\
    \end{cases}
\label{eqn:problem_spec}
\end{equation}
where $\odot$, $\boldsymbol{\tau}$ and $\texttt{clip}(\cdot)$  represent element-wise multiplication, trigger pattern and clipping operation, respectively. For each attack target $t_i$ there is the corresponding backdoor trigger $\tau_i$. 
Also, $\boldsymbol{\mathcal{M}}$ is the binary pruning mask with the same size of $\boldsymbol{\mathcal{W}}_\text{pt}$.





\textbf{Questions to be Answered.} As indicated by Eq. \ref{eqn:problem_spec}, a backdoored and pruned DNN model is jointly determined by the selection of the network pruning and backdoor trigger generation schemes. Considering the complicated interplay between these two schemes as well as their multiple design options, next we explore to answer the following three important questions towards developing high-performance pruned model-oriented backdoor attack.   

\underline{\textbf{Question \#1:}} \textit{What is the proper operational sequence when jointly performing network pruning and injecting backdoor triggers?}

\begin{figure}[ht]
\centering
\includegraphics[width=1.0\textwidth]{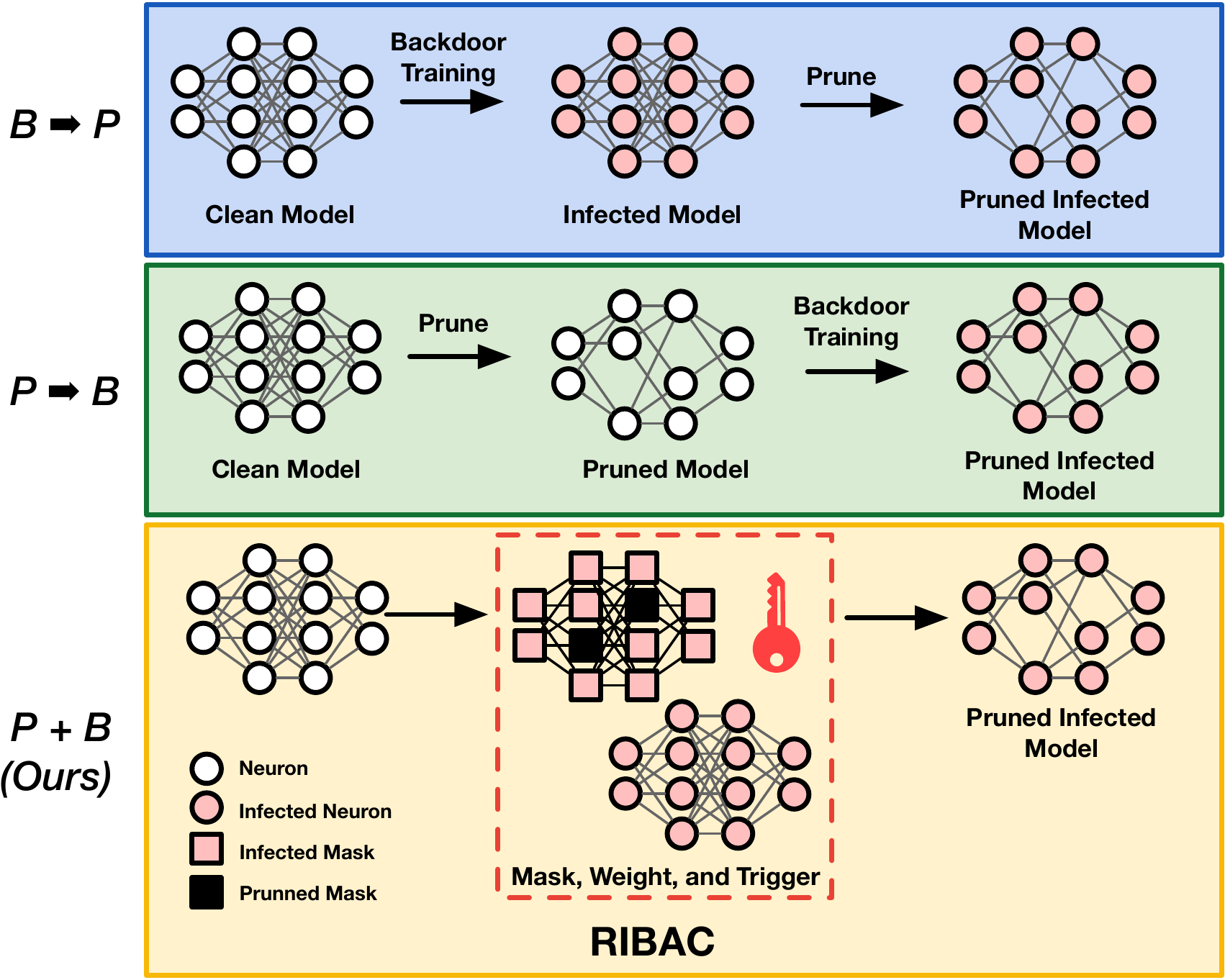}
\caption{Different operational sequences for obtaining backdoored pruned model. Given a clean model, \textit{\textbf{B $\rightarrow$ P}} first injects the backdoor and then performs pruning; \textit{\textbf{P $\rightarrow$ B}} first prunes the model and then performs backdoor training on the pruned networks; Our proposed \textit{\textbf{P + B}} learns the pruning masks, model weights and trigger patterns simultaneously.}
\label{figure:main_method}
\end{figure}

\underline{Analysis.} In general, imposing both network sparsity and backdoor triggers on the benign and uncompressed DNN models can be realized in different ways. The most straightforward solution is to perform network pruning and backdoor injection \textit{sequentially}. As illustrated in Figure \ref{figure:main_method}, we can either ``prune-then-inject" or ``inject-then-prune" to alter the original uncompressed and benign model to the desired compressed and backdoored one. For simplicity, we denote these two sequential schemes as \textit{\textbf{B $\rightarrow$ P}} and \textit{\textbf{P $\rightarrow$ B}}, where \textit{\textbf{B}}, \textit{\textbf{P}} represent the operation of injecting backdoor and pruning network, respectively. 

\begin{figure}[ht]
\centering
\includegraphics[width=1.0\textwidth]{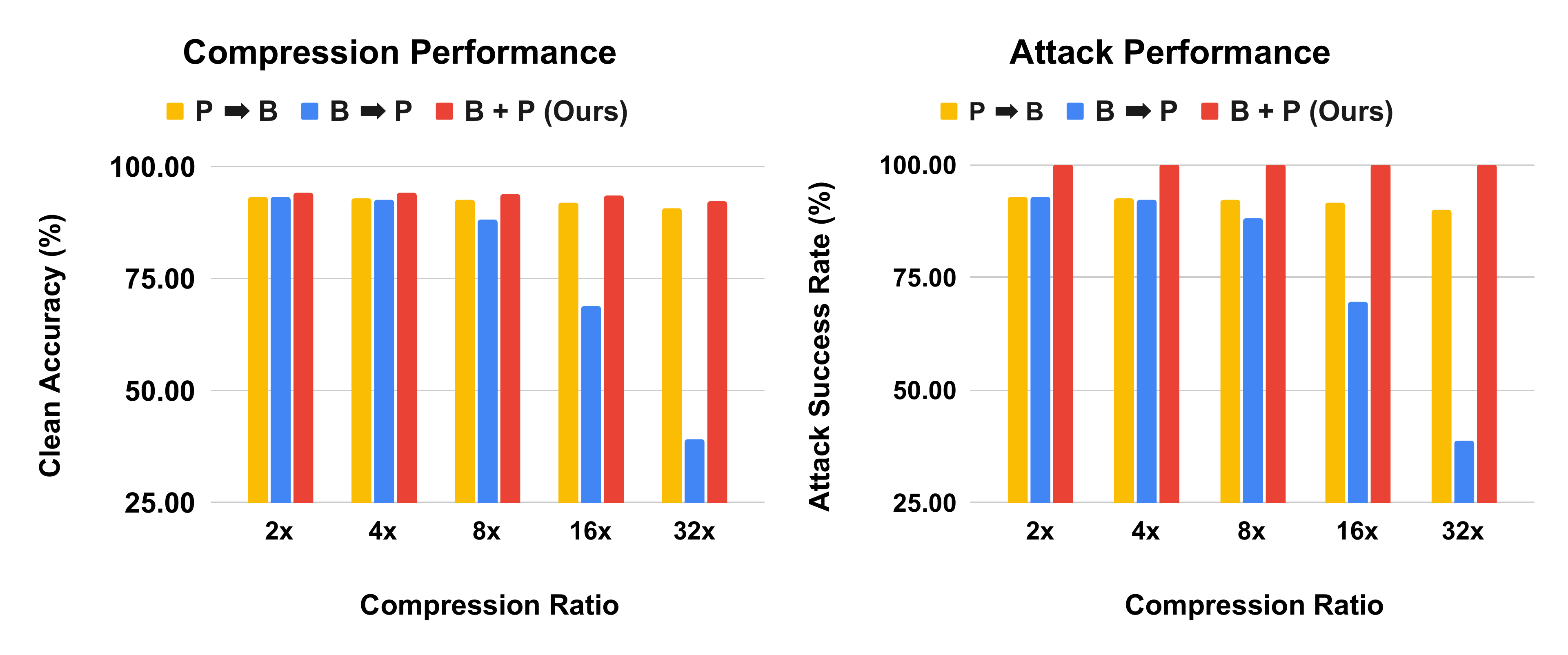}
\caption{Compression and Attack Performance of Preact ResNet-18 model on CIFAR-10 via using different operational sequences for pruning and backdoor injection. Here we adopt WaNet \cite{nguyen2021wanet} as the backdoor training method \textit{\textbf{B}} used in the \textit{\textbf{B $\rightarrow$ P}} and \textit{\textbf{P $\rightarrow$ B}} schemes.}
\label{figure:sequential_parallel}
\end{figure}

Evidently, the above two-stage schemes enjoy the benefit of convenient deployment since they can be easily implemented via simply combining the existing network pruning and backdoor attack approaches. However, we argue that they are not the ideal solutions when aiming for simultaneous high compression performance and attack performance. To be specific, because the current schemes for \textit{\textbf{B}} and \textit{\textbf{P}} are designed to optimize these two operations individually, the simple combination of these two locally optimal strategies does not necessarily bring globally optimal solution. For instance, as shown in Figure \ref{figure:sequential_parallel}, when aiming to launch backdoor attacks on a Preact ResNet-18 on CIFAR-10 dataset, both the \textit{\textbf{B $\rightarrow$ P}} and \textit{\textbf{P $\rightarrow$ B}} fail to achieve the satisfactory performance.

\underline{Our Proposal.} To perform joint network pruning and backdoor attack in an efficient way, we propose to adopt \textit{parallel} operational scheme (denoted as \textit{\textbf{P+B}}) for these two operations. To be specific, the compression-related design knobs, i.e., masking selection and weight update, and the attack-related design knobs, i.e., trigger pattern, will be determined together. To that end, Eq. \ref{eqn:problem_spec} is reformulated to the format with a unified objective function as follows:
\begin{align}
\begin{split}
\underset{\boldsymbol{\mathcal{W}}, \boldsymbol{\mathcal{M}}, \boldsymbol{\tau}}\min 
\mathcal{J} = 
\underset{\boldsymbol{\mathcal{W}}, \boldsymbol{\mathcal{M}}, \boldsymbol{\tau}}\min ~ & [ \underbrace{\mathcal{L}(\boldsymbol{\mathcal{W}}\odot\boldsymbol{\mathcal{M}}, \boldsymbol{x}, \boldsymbol{y})}_\text{clean data loss}
~ + ~ \beta \cdot \underbrace{\mathcal{L}(\boldsymbol{\mathcal{W}}\odot\boldsymbol{\mathcal{M}}, \texttt{clip}(\boldsymbol{x}+\boldsymbol{\tau}), \boldsymbol{t})}_\text{Trojan data loss} ],\\
& \text{s.t.} ~~ ||\boldsymbol{\mathcal{M}}||_0 \le s  ~~ \text{and} ~~ ||\boldsymbol{\tau}||_\infty \le \epsilon,
\end{split}
\label{eqn:question1}
\end{align}

\noindent where $s$ is the sparsity constraint, and $\epsilon$ the the trigger stealthiness constraint. Here the overall loss consists of the clean data loss and Trojan data loss, which measure the model compression performance (in term of clean accuracy) and attack performance (in term of attack success rate), respectively. With such unified loss function, the backdoor triggers $\boldsymbol{\tau}$, pruning masks $\boldsymbol{\mathcal{M}}$ and model weights $\boldsymbol{\mathcal{W}}$ can be now learned in an end-to-end and simultaneous way. Notice that $\beta$ is a hyper-parameter to control the balance between two loss terms.

\underline{\textbf{Question \#2:}} \textit{Which pruning criterion is more suitable for producing the backdoored sparse DNN models?}

\underline{Analysis.} Consider pruning serves as a key component of the compression-aware attack, the proper selection of the pruning mask $\boldsymbol{\mathcal{M}}$ is very critical. To date, weight magnitude-based pruning, which aims to remove the connections that have the least weights, is the most popular pruning method used in practice. In particular, several prior works \cite{ye2019adversarial,sehwag2019towards} that co-explore the sparsity and adversarial robustness of DNN models are also built on this pruning strategy.

\begin{figure}[ht]
\centering
\includegraphics[width=1.0\textwidth]{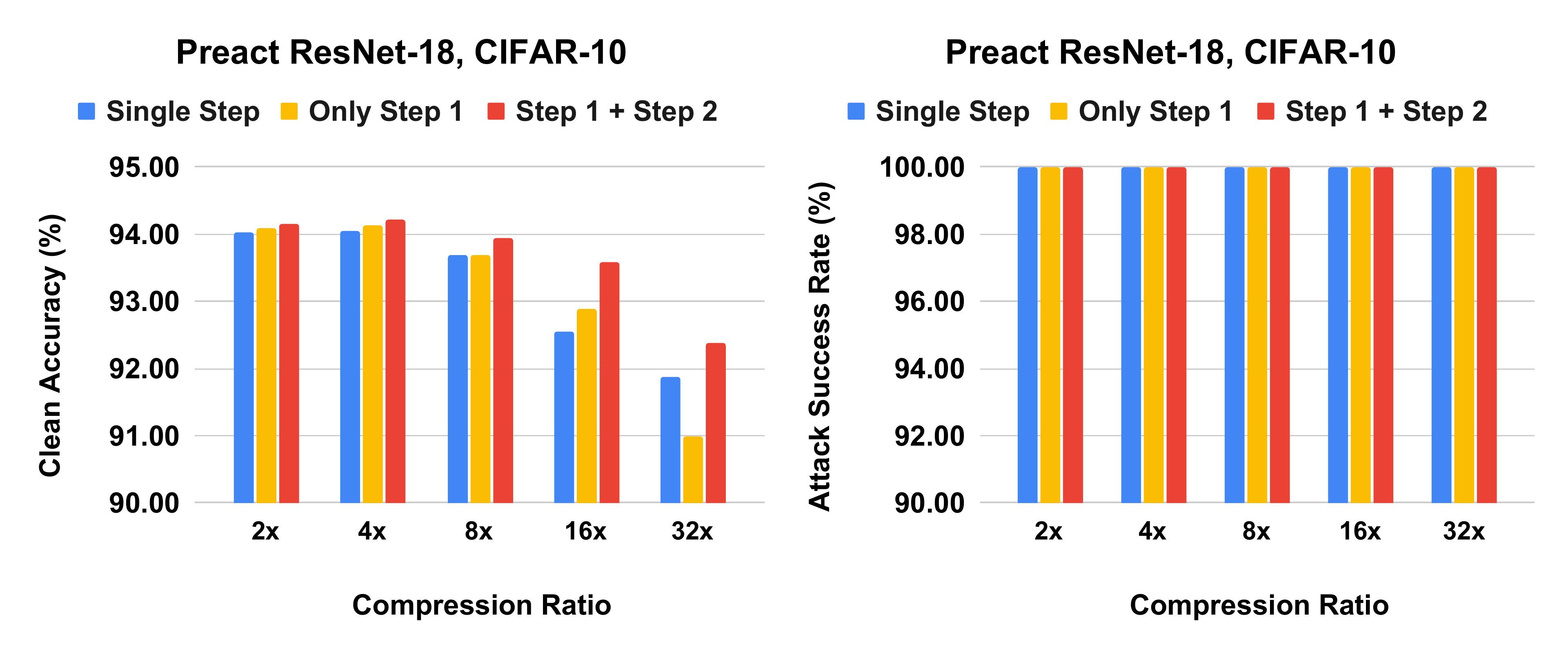}
\caption{Compression and Attack Performance of Preact ResNet-18 model on CIFAR-10 via using different schemes for training weights $\boldsymbol{\mathcal{W}}$, triggers $\boldsymbol{\tau}$ and importance scores $\boldsymbol{\mathcal{S}}$. \textbf{Single Step} means to train all of them simultaneously.  \textbf{Only Step 1} means to only train $\boldsymbol{\tau}$ and $\boldsymbol{\mathcal{S}}$. \textbf{Step 1 + Step 2} means to first train $\boldsymbol{\tau}$ and $\boldsymbol{\mathcal{S}}$, and then train $\boldsymbol{\tau}$ and $\boldsymbol{\mathcal{W}}$.}
\label{figure:step1_step2}
\end{figure}

However, we argue that the weight magnitude-based pruning is not the ideal solution towards producing a backdoored pruned model. Recall that the design philosophy for this pruning criterion is that the smaller weights intend to exhibit less importance. Although this assumption heuristically works when the overall task focuses on improving compression performance, it does not hold if other requirement, such as achieving high attack performance, needs to be satisfied. More specifically, the weights with less magnitudes does not necessarily mean that they are less important for the vulnerability of the model with the presence of backdoor attack. Consequently, if a DNN model is pruned via such pruning criterion ignoring the impacts on vulnerability, the resulting backdoor attack performance is likely to be very limited.

\underline{Our Proposal.} To address this issue, we propose to perform pruning in an attack-aware way. To that end, we choose to apply the philosophy of importance score \cite{ramanujan2020s} to the pruning process. More specifically, the trainable importance score, which measures the impact of the specific weight for the attack performance, is assigned to each neuron. Assume that $\boldsymbol{\mathcal{S}}$ be the set of importance scores of the weights, and let $m_i$ and $s_i$ be the $i^{th}$ element in the flatten $\boldsymbol{\mathcal{M}}$ and $\boldsymbol{\mathcal{S}}$, respectively. Then in the forward pass the mask $\boldsymbol{\mathcal{M}} = h(\boldsymbol{\mathcal{S}})$ is generated as:
\begin{equation}
    m_i = h(s_i) =
    \begin{cases}
      1 & \text{if } s_i \in \texttt{topK(}\boldsymbol{\mathcal{S}},k, l \texttt{)}, \\
      0 & \text{otherwise}
    \end{cases} 
\label{eqn:score_to_mask}
\end{equation}
where \texttt{topK$(\cdot, \cdot, \cdot)$} is the function that returns top $k\%$ highest score in layer $l$. During the training $\boldsymbol{\mathcal{S}}$ can be then updated with learning rate $\alpha_1$ as:
\begin{equation}
    \boldsymbol{\mathcal{S}} \gets \boldsymbol{\mathcal{S}} -  \alpha_1 \cdot \nabla_{\boldsymbol{\mathcal{S}}} [ \mathcal{J}(\boldsymbol{\mathcal{W}}, h(\boldsymbol{\mathcal{S}}), \boldsymbol{\tau}, \boldsymbol{x}, \boldsymbol{y})].
\label{eqn:update_s}
\end{equation}


\underline{\textbf{Question \#3:}} \textit{What is the proper learning scheme to perform the end-to-end training on pruning masks, trigger patterns and model weights?}

\underline{Analysis.} Eq. \ref{eqn:question1} shows that injecting backdoor trigger to the pruned model can be interpreted as the joint learning of masks, triggers and weights. To that end, a straightforward method is to directly optimize the unified loss described in Eq. \ref{eqn:question1}. However, this strategy is not an ideal solution because it ignores the complicated interplay among these three learnable objectives. For instance, the efforts for updating weights and masks may have opposite impacts on the compression performance, which may also further affect the attack performance. As illustrated in Figure \ref{figure:step1_step2}, such direct optimization strategy does not bring satisfied performance on compression aspect. 

\underline{Our Proposal.} To properly learn the suitable masks, triggers and weights to maximize the compression and attack performance, we propose to learn the masks and weights in two separate steps with always keeping the update of triggers. This idea is build on a key observation. As shown in Figure \ref{figure:step1_step2}, when we only train the importance score and trigger pattern (\textbf{Only Step 1}), even the weights are frozen to the initial values, very high attack success rate can already be obtained with slightly dropped clean accuracy. An intuitive explanation for this phenomenon is that since the initialization of weights inherit from the pre-trained model, as long as the masks and triggers are properly trained, the drop of clean accuracy is not very significant because of the existence of clean data loss in the overall loss (Eq. \ref{eqn:question1}). Motivated by this observation, we can first focus on learning the masks and triggers to achieve the desired attack performance, and then "fine-tune" the weights to further improve the compression performance. In general, this two-step scheme can be described as follows:


\begin{align}
\mathbf{Step-1:} & ~ \underset{\boldsymbol{\mathcal{S}}, \boldsymbol{\mathcal{\tau}}}\min ~\mathcal{L}(\boldsymbol{\mathcal{W}}_\text{pt} \odot h(\boldsymbol{\mathcal{S}}), \boldsymbol{x}, \boldsymbol{y})
+ \beta \cdot \mathcal{L}(\boldsymbol{\mathcal{W}}_\text{pt} \odot h(\boldsymbol{\mathcal{S}}), \texttt{clip}(\boldsymbol{x}+\boldsymbol{\tau}), \boldsymbol{t}), \\
\mathbf{Step-2:} & ~ \underset{\boldsymbol{\mathcal{W}}, \boldsymbol{\mathcal{\tau}}}\min \mathcal{L}(\boldsymbol{\mathcal{W}}\odot\boldsymbol{\mathcal{M}}, \boldsymbol{x}, \boldsymbol{y})
+ \beta \cdot \mathcal{L}(\boldsymbol{\mathcal{W}}\odot\boldsymbol{\mathcal{M}}, \texttt{clip}(\boldsymbol{x}+\boldsymbol{\tau}), \boldsymbol{t}).
\end{align}

\textbf{The Overall Algorithm.} Built upon the above analysis and proposals, we then integrate them together and develop the overall algorithm for training a pruned model to achieve simultaneous high compression performance and backdoor attack performance. The details of this procedure are described in Algorithm 1.

\begin{algorithm}[t]
\SetAlgorithmName{Alg}{}{}
\textbf{Input:} {Pre-trained model $\boldsymbol{\mathcal{W}}_\text{pt}$, sparsity $s$, clean images $\boldsymbol{x}$, labels $\boldsymbol{y}$, targets $\boldsymbol{t}$}, learning rates $\alpha_1, \alpha_2, \alpha_3$, balancing factor $\beta$.\\
\textbf{Output:} {Fine-tuned backdoored sparse model $\boldsymbol{\mathcal{W}}_\text{ft}$, optimized triggers $\boldsymbol{\tau}$}. \\
\
$\boldsymbol{\mathcal{S}} \gets \boldsymbol{\mathcal{W}}_\text{pt}$; $\boldsymbol{\tau} \gets \texttt{random}(\boldsymbol{x} \texttt{.shape})$ \textcolor{brown}{\textit{$\triangleright$ initialize scores and triggers.}} \\

\For ({~~~\textcolor{brown}{\textit{$\triangleright$ Step \#1. Optimize masks and triggers.}}}) {($x_i$, $y_i$, $t_i$) in ($\boldsymbol{x}$, $\boldsymbol{y}$, $\boldsymbol{t}$)}{
$\boldsymbol{\mathcal{M}} \gets \texttt{generate\_mask}(\boldsymbol{\mathcal{S}}, 1-s)$  \textcolor{brown}{\textit{$\triangleright$ via Equation (\ref{eqn:score_to_mask})}.}  \\
$\hat{y}_\text{clean}, ~ \hat{y}_\text{bd} \gets f_{\boldsymbol{\mathcal{W}}_\text{pt}\odot\boldsymbol{\mathcal{M}}}(x_i), ~ f_{\boldsymbol{\mathcal{W}}_\text{pt} \odot \boldsymbol{\mathcal{M}}}(\texttt{clip}(x_i + \boldsymbol{\tau}))$ \textcolor{brown}{\textit{$\triangleright$ forward pass.}}\\
$\mathcal{J} = \texttt{cross\_entropy}(\hat{y}_\text{clean}, y_i) + \beta \cdot \texttt{cross\_entropy}(\hat{y}_\text{bd}, t_i)$  \\
$\boldsymbol{\mathcal{S}} \gets \boldsymbol{\mathcal{S}} -  \alpha_1 \cdot \nabla_{\boldsymbol{\mathcal{S}}} [\mathcal{J}]$ \textcolor{brown}{\textit{$\triangleright$ update scores via Equation (\ref{eqn:update_s})}.}\\
$\boldsymbol{\mathcal{\tau}} \gets \Pi_\epsilon(\boldsymbol{\mathcal{\tau}} -  \alpha_2 \cdot \nabla_{\boldsymbol{\mathcal{\tau}}} [\mathcal{J}])$ \textcolor{brown}{\textit{$\triangleright$ update triggers using projected SGD}}.}

$\boldsymbol{\mathcal{W}} \gets \boldsymbol{\mathcal{W}}_\text{pt} $ \textcolor{brown}{\textit{$\triangleright$ Load pre-trained weight for fine-tuning.}}

\For ({~~~\textcolor{brown}{\textit{$\triangleright$ Step \#2. Optimize weights and triggers.}}}) {($x_i$, $y_i$, $t_i$) in ($\boldsymbol{x}$, $\boldsymbol{y}$, $\boldsymbol{t}$)}{
$\hat{y}_\text{clean}, ~ \hat{y}_\text{bd} \gets f_{\boldsymbol{\mathcal{W}}\odot\boldsymbol{\mathcal{M}}}(x_i), ~ f_{\boldsymbol{\mathcal{W}}\odot\boldsymbol{\mathcal{M}}}(\texttt{clip}(x_i + \boldsymbol{\tau}))$ \textcolor{brown}{\textit{$\triangleright$ forward pass.}}\\
$\mathcal{J} = \texttt{cross\_entropy}(\hat{y}_\text{clean}, y_i) + \beta \cdot \texttt{cross\_entropy}(\hat{y}_\text{bd}, t_i)$  \\
$\boldsymbol{\mathcal{W}} \gets \boldsymbol{\mathcal{W}} -  \alpha_3 \cdot \nabla_{\boldsymbol{\mathcal{W}}} [\mathcal{J}]$ \textcolor{brown}{\textit{$\triangleright$ update weight using SGD.}}\\
$\boldsymbol{\mathcal{\tau}} \gets \Pi_\epsilon(\boldsymbol{\mathcal{\tau}} -  \alpha_2 \cdot \nabla_{\boldsymbol{\mathcal{\tau}}} [\mathcal{J}])$ \textcolor{brown}{\textit{$\triangleright$ update triggers using projected SGD.}}}
$\boldsymbol{\mathcal{W}}_\text{ft} \gets \boldsymbol{\mathcal{W}}\odot\boldsymbol{\mathcal{M}} $ \textcolor{brown}{\textit{$\triangleright$ finalize the weight.}}

\caption{The Procedure of RIBAC Algorithm}
\label{algorithm1}
\end{algorithm}
 


\section{Experiments}

\subsection{Experiment Setup}
\textbf{Datasets and Models.} Following the prior works WaNet \cite{nguyen2021wanet}, LIRA \cite{doan2021lira} and WB \cite{doan2021backdoor}, we evaluate our method on four commonly used datasets for backdoor attacks: CIFAR-10 \cite{krizhevsky2009learning}, GTSRB \cite{stallkamp2012man}, CelebA \cite{liu2015deep}, and Tiny ImageNet \cite{le2015tiny}. We select Pre-Activate ResNet-18 \cite{he2016deep} for evaluation on CIFAR-10 and GTSRB datasets, and ResNet-18 \cite{he2016deep} for evaluation on CelebA and Tiny ImageNet datasets.

\textbf{Hyperparameter and Attack Setting.} We train the models for 60 epochs via using Adam optimizer with the learning rate of 0.0003. All the experiments are performed using Pytorch on Nvidia RTX 3090 GPU. To generate the target classes $\boldsymbol{t}$ for backdoor attacks, we adopt the two common all-to-one and all-to-all settings. For all-to-one configuration, we choose the first class as our target: $t_i = 0 ~ \forall ~ i$; for all-to-all configuration, the targets are the correct labels offset by 1: $t_i = y_i + 1 ~ \texttt{mod} ~ c ~ \forall i$, where $c$ is the number of classes. To ensure the stealthiness of our triggers $\boldsymbol{\tau}$, we use the operation $\Pi_\epsilon$ to clip the values of $\boldsymbol{\tau}$ that are outside the limit of $\epsilon = 4/255$.

\subsection{Attack Performance and Compression Performance}

\textbf{Comparison with Simple Combination of Pruning \& Backdoor Injection.} We compared the performance of RIBAC with other alternatives for obtaining the backdoored pruned model. Here we design three baseline methods: 1) Randomly initialize a sparse model, then train it using the state-of-the-art WaNet backdoor training \cite{nguyen2021wanet}; 2) prune a clean pre-trained network, then train it using WaNet backdoor training; 3) train a full-size model using WaNet backdoor training, then prune the model to achieve the target compression ratio.
As shown in Table \ref{table:compare_with_simple_combination}, all three baseline methods fail to achieve the satisfied performance. On the other hand, RIBAC can consistently achieve high clean accuracy and high attack success rate even at high compression ratio. In particular RIBAC can achieve up to $46.22\%$ attack success rate increase with 32$\times$ compression ratio on Tiny ImageNet dataset.

\begin{table}[t]
\setlength{\tabcolsep}{6pt}
\centering
\caption{Simple combination of pruning and backdoor injection versus RIBAC with respect to clean accuracy / attack success rate. C.R. means compression ratio.}
\scalebox{0.9}{
\begin{tabular}{
>{\columncolor[HTML]{FFFFFF}}c 
>{\columncolor[HTML]{FFFFFF}}c 
>{\columncolor[HTML]{FFFFFF}}c 
>{\columncolor[HTML]{FFFFFF}}c 
>{\columncolor[HTML]{EFEFEF}}c }
\toprule
\textbf{C.R.} & \textbf{\begin{tabular}[c]{@{}c@{}}  P $\rightarrow$ B \\ (Random Init.)\end{tabular}} & \textbf{\begin{tabular}[c]{@{}c@{}} P $\rightarrow$ B \\ (Clean Pre-trained)\end{tabular}} & \textbf{B $\rightarrow$ P} & \textbf{\begin{tabular}[c]{@{}c@{}}P + B\\ (RIBAC)\end{tabular}} \\ \midrule
\multicolumn{5}{c}{\cellcolor[HTML]{FFFFFF}\textit{\textbf{Preact ResNet-18 on CIFAR-10 dataset}}} \\
$2\times$ & 93.97 / 93.63 & 93.45 / 93.03 & 93.24 / 92.89 & \textbf{94.16 / 100.00} \\
$4\times$& 94.18 / 93.91 & 93.11 / 92.55 & 92.56 / 92.24 & \textbf{94.22 / 100.00} \\
$8\times$ & 93.29 / 92.95 & 92.73 / 92.22 & 88.33 / 88.23 & \textbf{93.94 / 100.00} \\
$16\times$ & 92.53 / 92.18 & 92.21 / 91.57 & 68.95 / 69.50 & \textbf{93.58 / 100.00} \\
$32\times$ & 89.25 / 88.59 & 90.90 / 89.99 & 39.14 / 38.93 & \textbf{92.39 / 100.00} \\
\multicolumn{5}{c}{\cellcolor[HTML]{FFFFFF}\textit{\textbf{ResNet-18 on CelebA dataset}}} \\
$2\times$ & 79.67 / 79.61 & 79.32 / 79.34 & 77.00 / 76.95 & \textbf{81.87 / 100.00} \\
$4\times$ & 79.54 / 79.50 & 79.26 / 79.17 & 75.43 / 75.40 & \textbf{81.52 / 100.00} \\
$8\times$ & 79.83 / 79.74 & 79.07 / 79.06 & 51.38 / 51.63 & \textbf{81.57 / 100.00} \\
$16\times$ & 79.75 / 79.59 & 78.36 / 78.25 & 27.16 / 27.16 & \textbf{81.68 / 100.00} \\
$32\times$ & 79.69 / 79.75 & 79.28 / 78.77 & 27.16 / 27.16 & \textbf{81.68 / 100.00} \\
\multicolumn{5}{c}{\cellcolor[HTML]{FFFFFF}\textit{\textbf{ResNet-18 on Tiny ImageNet dataset}}} \\
$2\times$ & 61.85 / 60.97 & 58.83 / 58.25 & 59.84 / 59.29 & \textbf{60.19 / 99.31} \\
$4\times$ & 60.72 / 60.04 & 57.46 / 56.40 & 40.81 / 40.56 & \textbf{60.76 / 99.64} \\
$8\times$ & 59.04 / 58.64 & 57.04 / 56.37 & 1.64 / 1.53 & \textbf{60.41 / 99.07} \\
$16\times$ & 56.13 / 54.51 & 56.28 / 55.19 & 0.50 / 0.50 & \textbf{59.11 / 99.40} \\
$32\times$ & 50.16 / 49.28 & 54.28 / 53.06 & 0.50 / 0.50 & \textbf{54.99 / 99.28} \\
\bottomrule
\end{tabular}
}
\label{table:compare_with_simple_combination}
\end{table}

\begin{table}[t]
\setlength{\tabcolsep}{8pt}
\centering
\caption{RIBAC vs Pruning-only Approach. Here for pruning-only baseline only clean accuracy is reported, and the clean accuracy / attack success rate is listed for RIBAC.}
\scalebox{0.9}{
\begin{tabular}{
>{\columncolor[HTML]{FFFFFF}}c 
>{\columncolor[HTML]{FFFFFF}}c 
>{\columncolor[HTML]{FFFFFF}}c 
>{\columncolor[HTML]{EFEFEF}}c l
>{\columncolor[HTML]{EFEFEF}}c }
\toprule
\textbf{C.R} & \textbf{\begin{tabular}[c]{@{}c@{}}L1\\ Prune\end{tabular}} & \textbf{\begin{tabular}[c]{@{}c@{}}Important Score\\ Prune\end{tabular}} & \textbf{\begin{tabular}[c]{@{}c@{}}RIBAC\\ (all-to-one)\end{tabular}} &  & \textbf{\begin{tabular}[c]{@{}c@{}}RIBAC\\ (all-to-all)\end{tabular}} \\ \midrule
\multicolumn{6}{c}{\cellcolor[HTML]{FFFFFF}\textit{\textbf{Preact ResNet-18 (Pretrain 94.61) on CIFAR-10 dataset}}} \\
$2\times$ & 94.51 & 94.61 & 94.35 / 100.00 &  & 94.16 / 100.00 \\
$4\times$ & 94.74 & 94.60 & 94.57 / 100.00 &  & 94.22 / 100.00 \\
$8\times$ & 94.86 & 94.31 & 94.36 / 100.00 &  & 93.94 / 100.00 \\
$16\times$ & 94.01 & 94.07 & 94.29 / 100.00 &  & 93.58 / 100.00 \\
$32\times$ & 91.51 & 93.05 & 91.77 / 100.00 &  & 92.39 / 100.00 \\
\multicolumn{6}{c}{\cellcolor[HTML]{FFFFFF}\textit{\textbf{Preact ResNet-18 (Pretrain 99.07) on GTSRB dataset}}} \\
$2\times$ & 98.87 & 99.11 & 98.85 / 100.00 &  & 99.03 / 99.98 \\
$4\times$ & 98.86 & 98.50 & 98.48 / 100.00 &  & 98.96 / 99.97 \\
$8\times$ & 98.39 & 98.74 & 98.36 / 100.00 &  & 98.48 / 100.00 \\
$16\times$ & 98.86 & 98.65 & 98.80 / 100.00 &  & 98.00 / 99.02 \\
$32\times$ & 97.33 & 97.91 & 98.04 / 100.00 &  & 96.92 / 98.34 \\
\multicolumn{6}{c}{\cellcolor[HTML]{FFFFFF}\textit{\textbf{ResNet-18 (Pre-train 60.08) on Tiny ImageNet dataset}}} \\
$2\times$ & 60.70 & 61.05 & 60.45 / 99.98 &  & 60.19 / 99.31 \\
$4\times$ & 60.86 & 61.25 & 60.70 / 99.95 &  & 60.76 / 99.64 \\
$8\times$ & 60.19 & 61.40 & 60.48 / 99.70 &  & 60.41 / 99.07 \\
$16\times$ & 59.20 & 60.25 & 59.65 / 99.92 &  & 59.11 / 99.40 \\
$32\times$ & 55.64 & 53.42 & 53.98 / 99.72 &  & 54.99 / 99.28 \\
\bottomrule
\end{tabular}}
\label{table:clean_baselines}
\end{table}

\textbf{Comparison with Standard Pruning Methods on Clean Accuracy.}
We also compare the compression performance of RIBAC with two standard pruning approaches: $L_1$ global pruning  \cite{han2015learning} and importance score-based pruning \cite{ramanujan2020s}. As reported in Table \ref{table:clean_baselines}, RIBAC can achieve the similar clean accuracy to the standard pruning approach with different pruning ratio, and meanwhile RIBAC can still achieve very high attack success rate. In other words, \ul{RIBAC does not trade compression performance for attack performance.} 

\begin{table}[t]
\setlength{\tabcolsep}{4pt}
\centering
\caption{Comparison with different backdoor attack methods with respect to clean accuracy / attack success rate. C.R. means compression ratio.}
\scalebox{0.9}{
\begin{tabular}{lcccc}
\rowcolor[HTML]{FFFFFF}
\toprule
\multicolumn{1}{c}{\cellcolor[HTML]{FFFFFF}\textbf{Method}} & \textbf{C.R.} & \textbf{\begin{tabular}[c]{@{}c@{}}Preact ResNet-18\\ CIFAR-10\end{tabular}} & \textbf{\begin{tabular}[c]{@{}c@{}}Preact ResNet-18\\ GTSRB\end{tabular}} & \textbf{\begin{tabular}[c]{@{}c@{}}ResNet-18\\ Tiny ImageNet\end{tabular}} \\
\rowcolor[HTML]{FFFFFF} \midrule
\multicolumn{5}{c}{\cellcolor[HTML]{FFFFFF}\textit{\textbf{All-to-one Backdoor Attacks}}} \\
\rowcolor[HTML]{FFFFFF}
WaNet\cite{nguyen2021wanet} & n/a & 94.15 /  99.55 & 98.97 / 98.78 & 57.00 / 99.00 \\
\rowcolor[HTML]{FFFFFF} 
LIRA\cite{doan2021lira}& n/a & 94.00 /  100.00 & 99.00 / 100.00 & 58.00 / 100.00 \\
\rowcolor[HTML]{FFFFFF} 
WB\cite{doan2021backdoor}& n/a & 94.00 /  99.00 & 99.00 / 99.00 & 57.00 / 100.00 \\
\rowcolor[HTML]{EFEFEF} 
RIBAC & $2\times$ & 94.35 / 100.00 & 98.85 / 100.00 & 60.45 / 99.98 \\
\rowcolor[HTML]{EFEFEF} 
RIBAC & $4\times$ & 94.57 / 100.00 & 98.48 / 100.00 & 60.70 / 99.95 \\
\rowcolor[HTML]{EFEFEF} 
RIBAC & $8\times$ & 94.36 / 100.00 & 98.36 / 100.00 & 60.48 / 99.70 \\
\rowcolor[HTML]{EFEFEF} 
RIBAC & $16\times$ & 94.29 / 100.00 & 98.80 / 100.00 & 59.65 / 99.92 \\
\rowcolor[HTML]{EFEFEF} 
RIBAC & $32\times$ & 91.77 / 100.00 & 98.04 / 100.00 & 53.98 / 99.72 \\
\rowcolor[HTML]{FFFFFF} 
\multicolumn{5}{c}{\cellcolor[HTML]{FFFFFF}\textit{\textbf{All-to-all Backdoor Attacks}}} \\
\rowcolor[HTML]{FFFFFF} 
WaNet\cite{nguyen2021wanet} & n/a & 94.00 /  93.00 & 99.00 / 98.00 & 58.00 / 58.00 \\
\rowcolor[HTML]{FFFFFF} 
LIRA\cite{doan2021lira}& n/a & 94.00 /  94.00 & 99.00 / 100.00 & 58.00 / 59.00 \\
\rowcolor[HTML]{FFFFFF} 
WB\cite{doan2021backdoor}& n/a & 94.00 /  94.00 & 99.00 / 98.00 & 58.00 / 58.00 \\
\rowcolor[HTML]{EFEFEF} 
RIBAC & $2\times$ & 94.16 / 100.00 & 99.03 / 99.98 & 60.19 / 99.31 \\
\rowcolor[HTML]{EFEFEF} 
RIBAC & $4\times$ & 94.22 / 100.00 & 98.96 / 99.97 & 60.76 / 99.64 \\
\rowcolor[HTML]{EFEFEF} 
RIBAC & $8\times$ & 93.94 / 100.00 & 98.48 / 100.00 & 60.41 / 99.07 \\
\rowcolor[HTML]{EFEFEF} 
RIBAC & $16\times$ & 93.58 / 100.00 & 98.00 / 99.02 & 59.11 / 99.40 \\
\rowcolor[HTML]{EFEFEF} 
RIBAC & $32\times$ & 92.39 / 100.00 & 96.92 / 98.34 & 54.99 / 99.28 \\
\bottomrule
\end{tabular}}
\label{table:sota}
\end{table}

\textbf{Comparison with State-of-the-art Backdoor Attack Methods.}
We also compare RIBAC with the state-of-the-art backdoor attacks approaches WaNet \cite{nguyen2021wanet}, LIRA \cite{doan2021lira}, WB \cite{doan2021backdoor}. Notice that these existing methods cannot compress models. As shown in Table \ref{table:sota}, on CIFAR-10 and GTSRB datasets, RIBAC can achieve very similar or higher clean accuracy and attack success rate while providing additional compression benefits. On Tiny ImageNet dataset RIBAC outperforms the state-of-the-art backdoor attack methods with up to $2.76\%$ clean accuracy and $40.64\%$ attack success rate increase.

\subsection{Performance Against Defense Methods}
To demonstrate the robustness and stealthiness of the backdoor attack enabled by our proposed RIBAC, we evaluate its performance against several state-of-the-art backdoor defense methods.

\begin{figure}[t]
\centering
\includegraphics[width=1.0\textwidth]{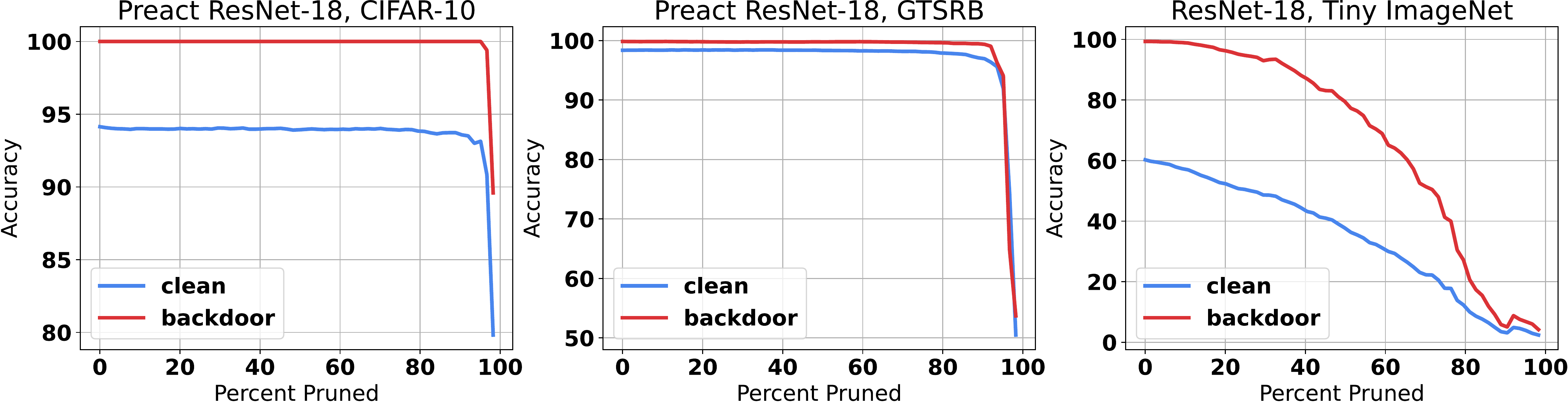}
\caption{Performance of RIBAC against Fine-Pruning.}
\label{figure:defense_fine_pruning}
\end{figure}

\textbf{Fine-Pruning \cite{liu2018fine}} argues that in a backdoored neural network there exist two groups of neurons that are associated with the clean images and backdoor triggers, respectively. Based on this assumption and with a small set of clean images, Fine-Pruning records the activation maps of the neurons in last convolution layer, and then gradually prunes these neurons based on activation magnitude to remove the backdoor. Figure \ref{figure:defense_fine_pruning} shows the performance of Fine-Pruning on the model generated by RIBAC. As the number of pruned neuron increases, both the clean accuracy and attack success rate gradually drop. However, the attack success rate of RIBAC is always higher than the clean accuracy, thereby making Fine-Pruning fail to mitigate the backdoor.

\begin{figure}[t]
\centering
\includegraphics[width=1.0\textwidth]{figures/strip.pdf}
\caption{Performance of RIBAC against STRIP.}
\label{figure:defense_strip}
\end{figure}

\begin{figure}[t]
\centering
\includegraphics[width=0.7\textwidth]{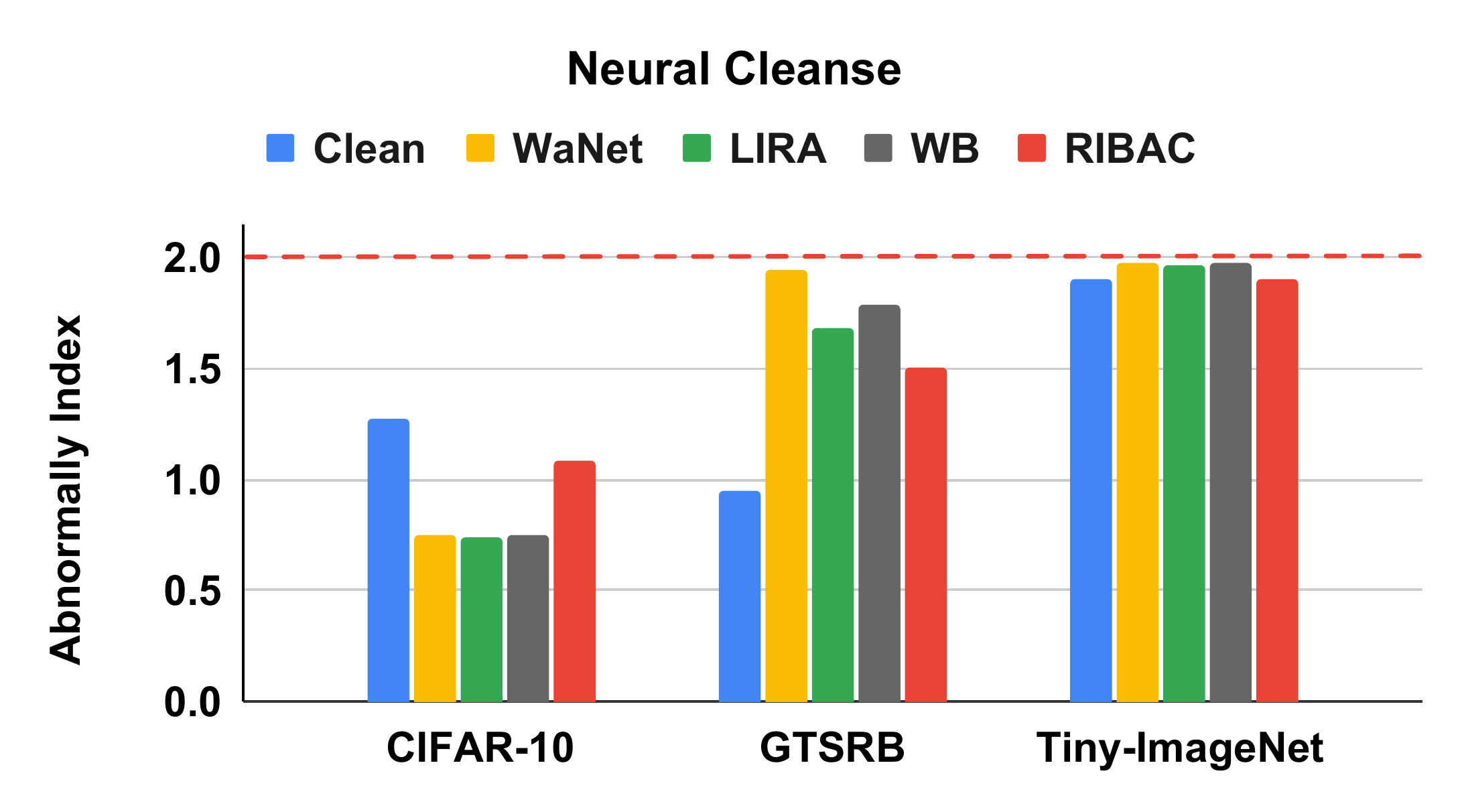}
\caption{Performance of RIBAC against Neural Cleanse.}
\label{figure:defense_neural_cleanse}
\end{figure}

\begin{figure}[t]
\centering
\includegraphics[width=1.0\textwidth]{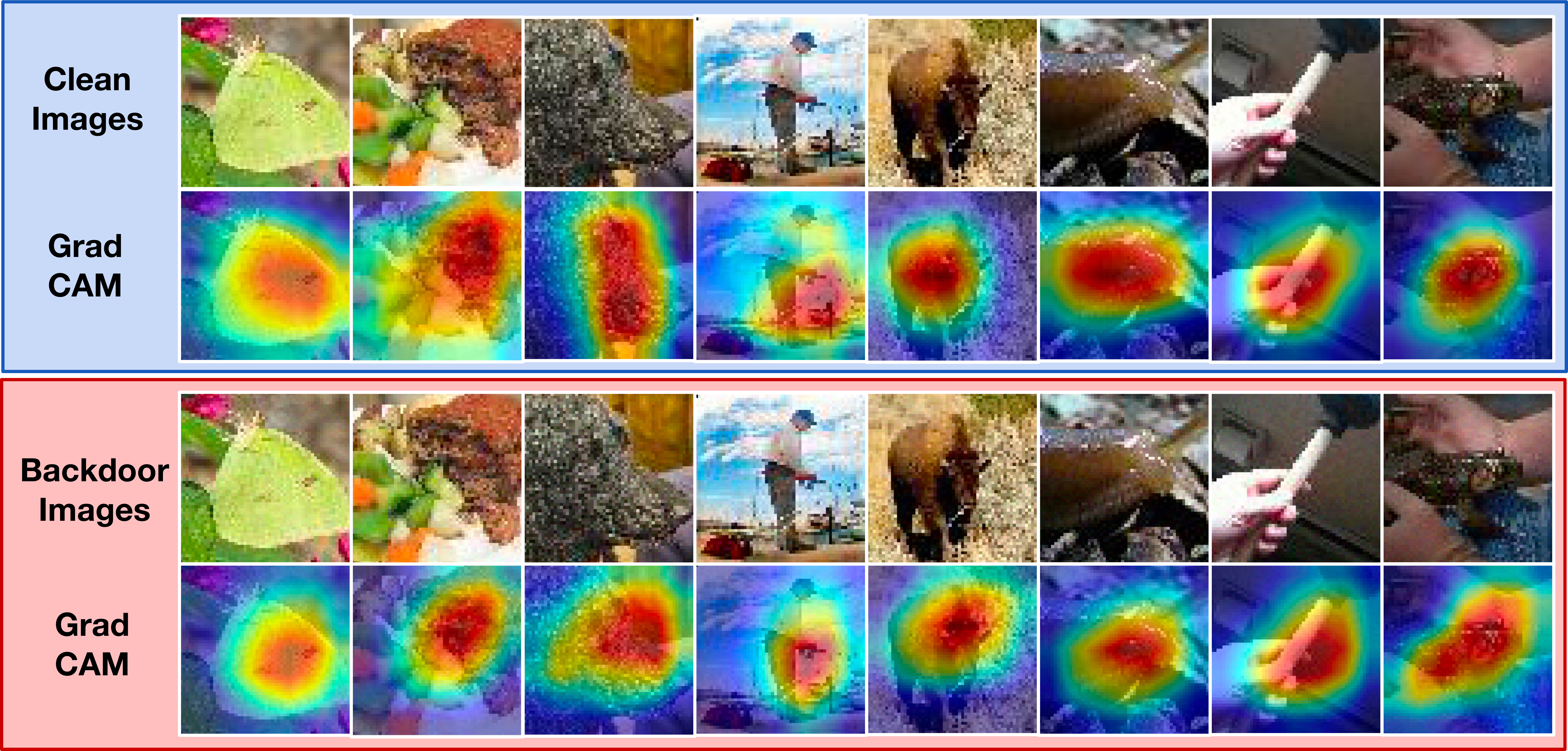}
\caption{Visualization of heatmap via GradCAM.}
\label{figure:defense_grad_cam}
\end{figure}

\textbf{STRIP \cite{gao2019strip}} focuses on analyzing the entropy of the prediction. Its key idea is to perform perturbation on the input image via using
a set of benign inputs from different classes. If the predictions of these perturbed inputs are persistent, which corresponds to low entropy, the potential presence of backdoor will be alarmed. For RIBAC, because the learned backdoor triggers are imperceptible and extremely stealthy ($||\boldsymbol{\tau}||_\infty \le 4 / 255)$, the perturbation operation adopted in STRIP effectively modifies the triggers, making our backdoored model behave like a clean model with similar entropy range (see Figure \ref{figure:defense_strip}).

\textbf{Neural Cleanse \cite{wang2019neural}} assumes that there exists patch-based pattern causing the misclassification. Based on this assumption, Neural Cleanse performs optimization to calculate the patch pattern that can altering the clean input to the target label. If a significant smaller pattern exists for any class label, a sign of potential backdoor will be alarmed. Such decision is quantified via using Abnormally Index with a threshold = $2.0$, which determines the existence of backdoor or not. As shown in Figure \ref{figure:defense_neural_cleanse}, our RIBAC passes all the Neural Cleanse tests across different datasets. For the test on CIFAR-10 and Tiny ImageNet, RIBAC can even achieve similar scores to the clean model.

\textbf{GradCAM \cite{selvaraju2017grad}}, as a method to visualize the network attention for the input image, can serve as an inspection tool to check the potential presence of backdoor. Figure \ref{figure:defense_grad_cam} shows the visualization of the network's attention for both clean and trigger-contained images. It is seen that the heat map of RIBAC looks similar to the one from clean model, thereby making it passes the GradCAM-based inspection.

\section{Conclusion}
In this paper, we propose RIBAC, a robust and imperceptible backdoor attack against compact DNN models. The proper trigger patterns, model weights and pruning masks are simultaneously learned in an efficient way. Experimental results across different datasets show that RIBAC attack exhibits high stealthiness, high robustness and high model efficiency. 

\subsubsection{Acknowledgement}
This work was partially supported by National Science Foundation under Grant
CNS2114220,
CCF1909963,
CCF2211163,
CNS2114161,
CCF-2000480,
CCF-2028858,
CNS-2120276,
and CNS-2145389.

\bibliographystyle{splncs04}
\bibliography{main_bib}
\end{document}


\pagestyle{headings}
\mainmatter
\def\ECCVSubNumber{3867}  

\title{Supplementary Material for ``RIBAC: Towards \ul{R}obust and \ul{I}mperceptible \ul{B}ackdoor \ul{A}ttack against \ul{C}ompact DNN''} 


\titlerunning{RIBAC}
%
\author{Huy Phan\inst{1} \and
Cong Shi\inst{1} \and
Yi Xie\inst{1} \and
Tianfang Zhang\inst{1} \and
Zhuohang Li\inst{2} \and
Tianming Zhao\inst{3} \and
Jian Liu\inst{2} \and
Yan Wang\inst{3} \and
Yingying Chen\inst{1} \and
Bo Yuan \inst{1}
}
\authorrunning{H. Phan et al.}
%
\institute{Rutgers University, New Jersey, USA \and
The University of Tennessee, Tennessee, USA \and Temple University, Pennsylvania, USA}

\maketitle

\section{Ablation study on Trigger Stealthiness.}

\begin{figure}[h]
\centering
\includegraphics[width=0.9\textwidth]{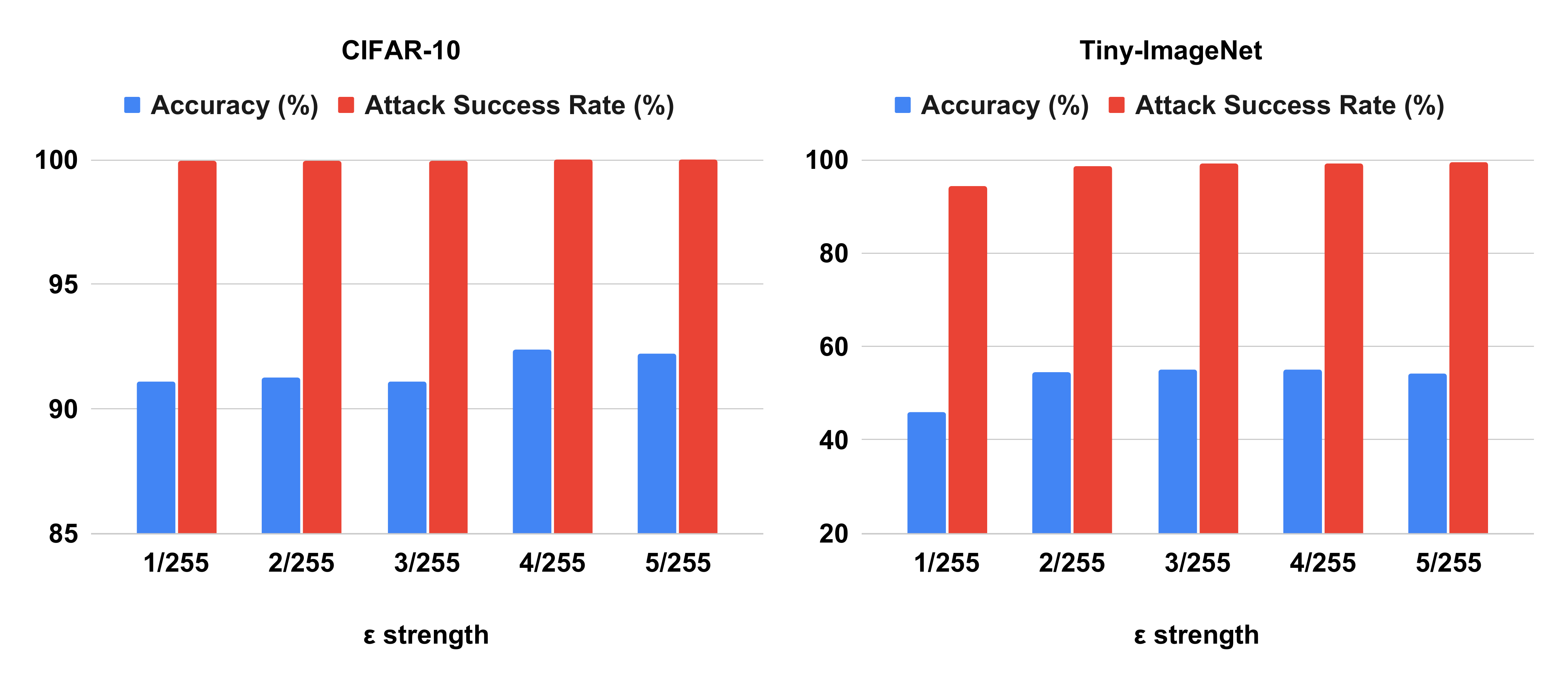}
\caption{Performance of RIBAC with varying trigger stealthiness ($\epsilon$) on CIFAR-10 and Tiny-ImageNet dataset.}
\vspace{-5mm}
\label{figure:trigger_stealthiness}
\end{figure}

We examine the effect of varying the maximum allowed perturbation $\epsilon$ on compression performance and attack performance. It is seen from Figure \ref{figure:trigger_stealthiness} that smaller values of $\epsilon$ (from $1/255$ to $3/255$), while can offer better stealthiness, suffer from degraded compression performance and attack performance. Higher value of $\epsilon$ ($5/255$) does not offer additional performance in both metrics. Hence, we believe that our default value of $\epsilon = 4/255$ gives a good balance between stealthiness, compression performance and attack performance.

\section{Ablation study on Number of Training Epochs.}
\begin{figure}[h]
\centering
\includegraphics[width=0.95\textwidth]{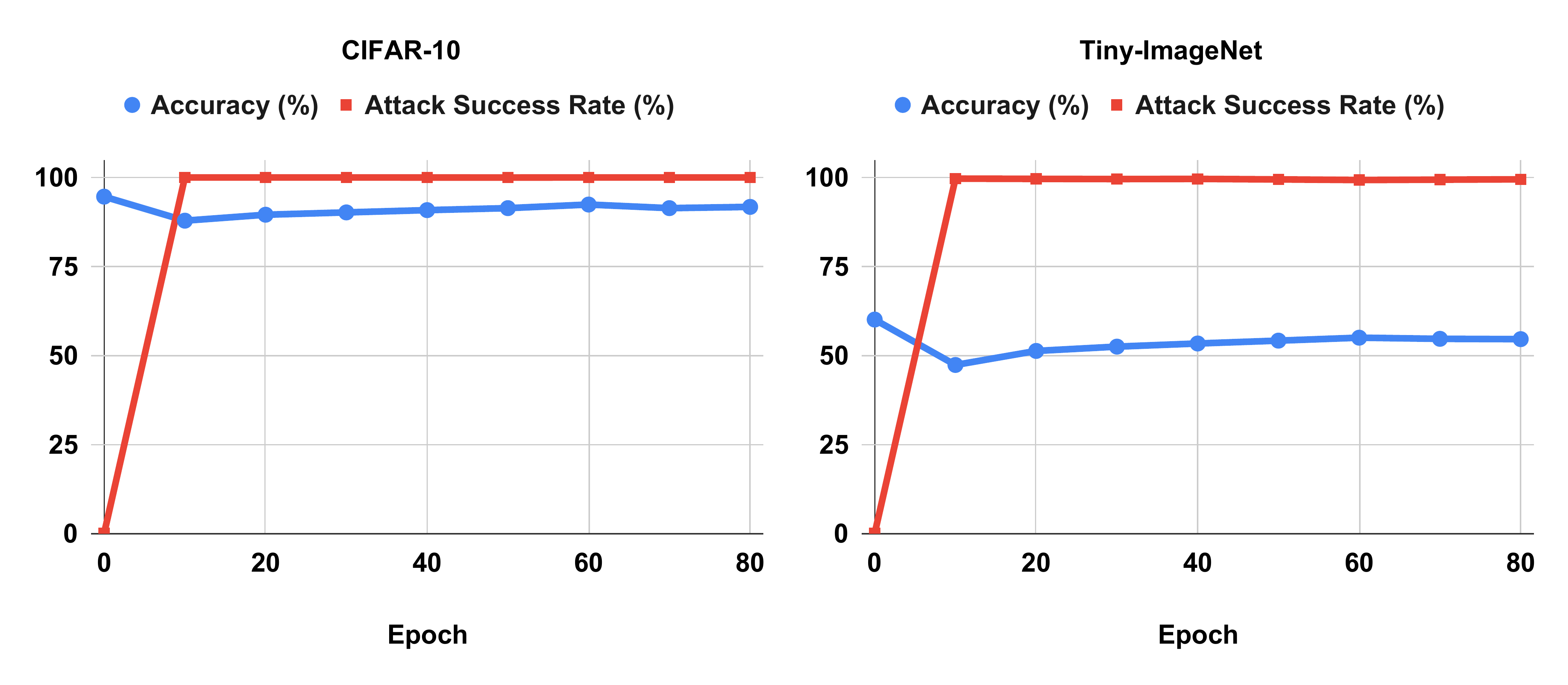}
\caption{Performance of RIBAC with varying number of training epochs on CIFAR-10 and Tiny-ImageNet dataset.}
\label{figure:training_epoch}
\end{figure}

We study the effect of changing the number of training epochs of $\mathbf{Step-1}$ in Eq. (6) and $\mathbf{Step-2}$ in Eq. (7). We can observe from Figure \ref{figure:training_epoch} that we can already achieve very high attack performance only by using a small number of training epochs. However, to recover the clean accuracy of the pre-trained models, more training epochs are needed. Since the clean accuracy stop increase after $60^{th}$ epoch, it is seen that our default value of using 60 training epochs for RIBAC offer a good balance between efficiency and performance. 

\section{Visual results of RIBAC backdoor images and triggers.}

\begin{figure}[h!]
\centering
\includegraphics[width=1.0\textwidth]{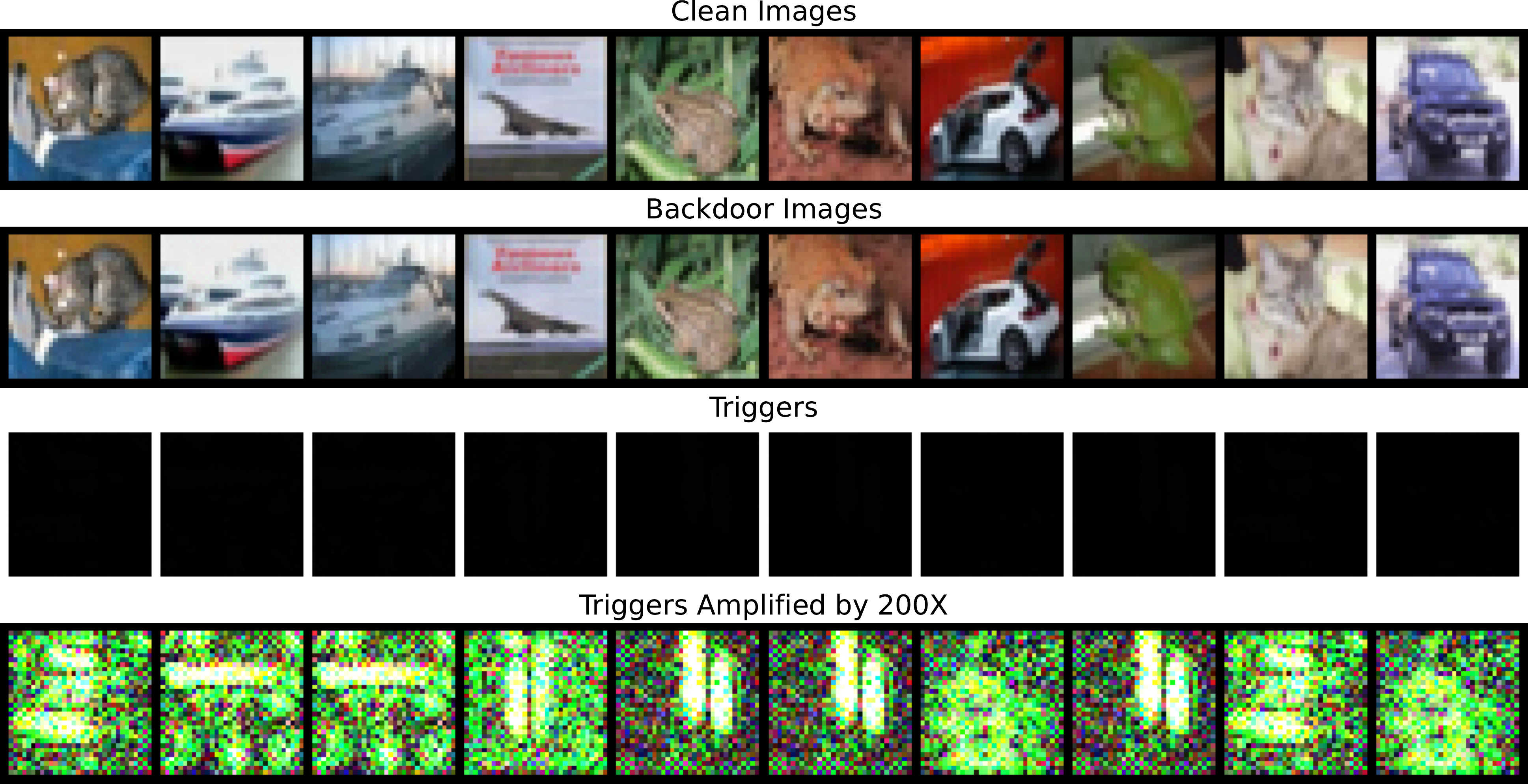}
\caption{Clean images, RIBAC backdoor images, and RIBAC amplified triggers on CIFAR-10 dataset.}
\label{figure:images_cifar10}
\end{figure}

\begin{figure}[h!]
\centering
\includegraphics[width=1.0\textwidth]{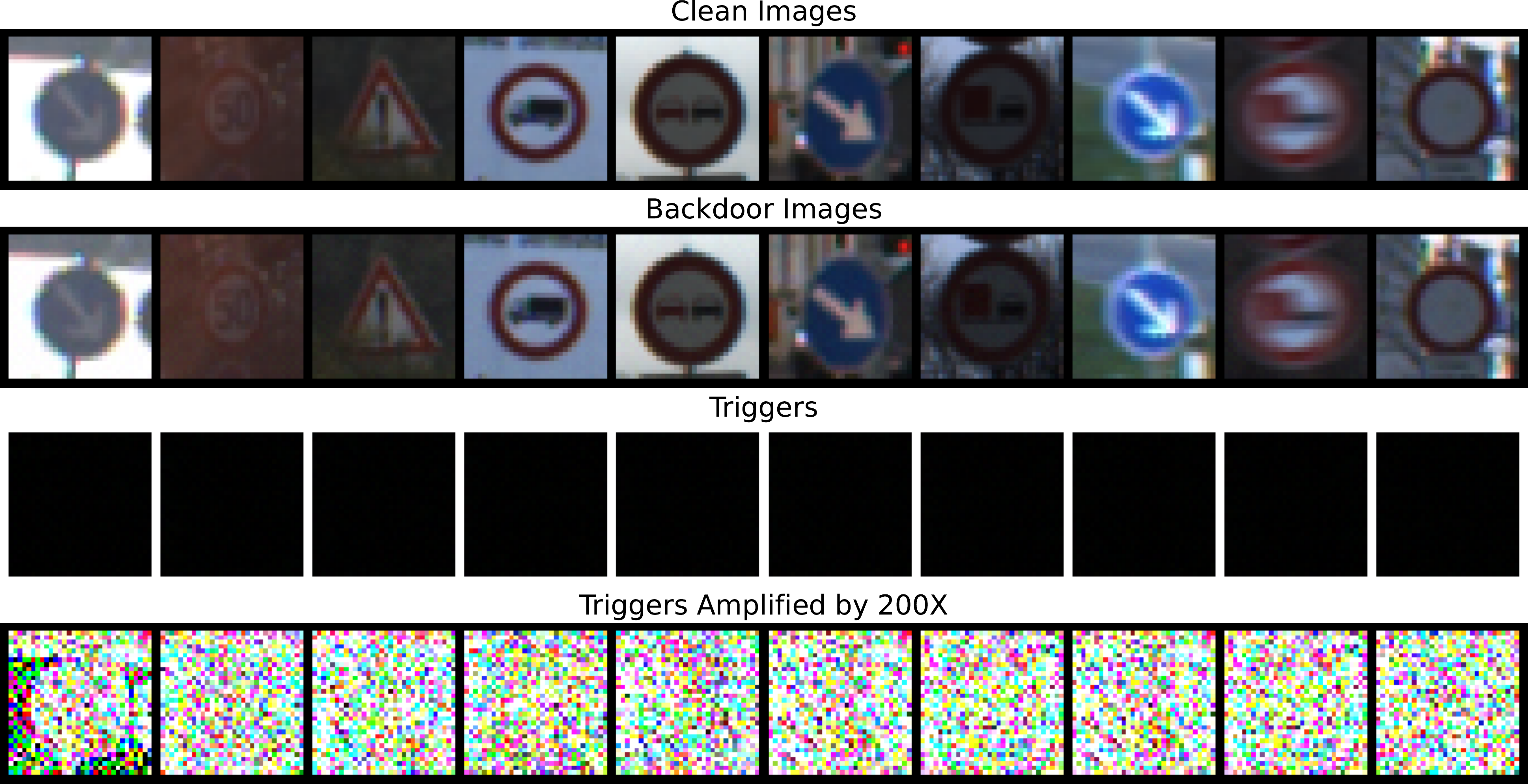}
\caption{Clean images, RIBAC backdoor images, and RIBAC amplified triggers on GTSRB dataset.}
\label{figure:images_gtsrb}
\end{figure}

\begin{figure}[h]
\centering
\includegraphics[width=1.0\textwidth]{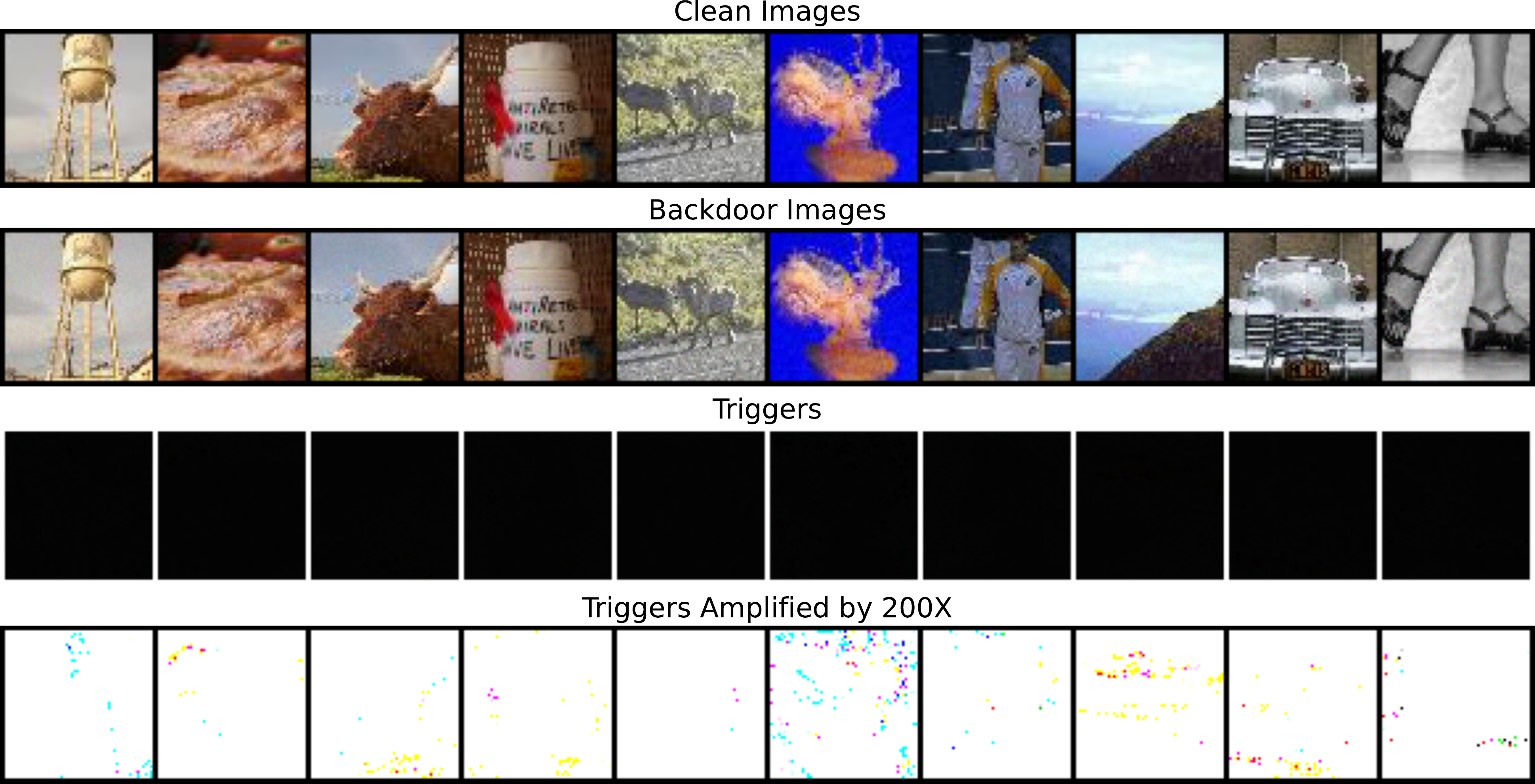}
\caption{Clean images, RIBAC backdoor images, and RIBAC amplified triggers on Tiny-ImageNet dataset.}
\label{figure:images_tiny_imagenet}
\end{figure}

To demonstrate the stealthiness of RIBAC backdoor images using different datasets, we show the clean images, backdoor images, and amplified triggers in Figure \ref{figure:images_cifar10}, Figure \ref{figure:images_gtsrb}, Figure  \ref{figure:images_tiny_imagenet}. It is seen that RIBAC backdoor images are visually indistinguishable from the clean images. 

